**Combined AFM and STM measurements of a silicene sheet grown on Ag(111) surface**

Zsolt Majzik[1], Mohamed Rachid Tchalala[2], Martin Švec[1], Prokop Hapala[1], Hanna Enriquez[2], Abdelkader Kara[3], Andrew J. Mayne[2], Gérald Dujardin[2], Pavel Jelínek[1] and Hamid Oughaddou[2,4]

[1]*Insitute of Physics, Academy of Sciences of the Czech Republic, Cukrovarnicka 10, Praha, 16200, Czech Republic*

[2]*Institut des Sciences Moléculaires d'Orsay, ISMO-CNRS, Bât. 210, Université Paris-Sud, F-91405 Orsay, France*

[3]*Department of Physics, University of Central Florida, Orlando, FL 32816, USA*

[4]*Département de Physique, Université de Cergy-Pontoise, F-95031 Cergy-Pontoise Cedex, France*

Abstract

In this Letter, we present the first non-contact atomic force microscopy (nc-AFM) of a silicene on silver (Ag) surface, obtained by combining non-contact atomic force microscopy (nc-AFM) and scanning tunneling microscopy (STM). STM images over large areas of silicene grown on Ag(111) surface show both (√13x√13)R13.9° and (4x4) superstructures. For the widely observed (4x4) structure, the nc-AFM topography shows an atomic-scale contrast inversion as the tip-surface distance is decreased. At the shortest tip-surface distance, the nc-AFM topography is very similar to the STM one. The observed structure in the nc-AFM topography is compatible with only one out of two silicon atoms being visible. This indicates unambiguously a strong buckling of the silicene honeycomb layer.

Advances in nanoscience involving low dimensional carbon based structures such as carbon nanotubes and graphene are attractive because of their interesting electronic properties.[1,2] Recently, much attention has been turned to the question of whether the silicon counterpart, namely silicene, can be fabricated and will it exhibit interesting properties. The main difference is that for carbon $sp^2$ hybridization is stable while silicon tends to prefer $sp^3$ hybridization. Theoretical investigations have shown that slightly buckled silicene has an intrinsic stability and presents electronic properties similar to those of graphene.[3-6] Experimentally, we have succeed in synthesizing silicene on both the Ag(110) and Ag(111) substrates.[7-10] On the Ag(110) surface, massively parallel silicene nano-ribbons (NRs) of 1.6 nm wide were obtained.[7,10] We have shown that their reactivity towards molecular oxygen is substantially less than that of silicon.[11] In addition, angle-resolved photoelectron spectroscopy (ARPES) measurements showed quantum confined electronic states of a one-dimensional character[12] with a dispersion along the length of the NRs, in the vicinity of the X-point of the Brillouin zone suggesting a behavior analogous to the Dirac cones of graphene.[12] On the Ag(111) surface, a continuous two dimensional (2D) sheet of silicene presenting a (2√3x2√3) R30° superstructure, has been observed by scanning tunneling microscopy (STM).[9] Following our pioneering work, several groups have successfully grown silicene on Ag(111) and reported the existence of different ordered phases[13-16] (2√3x2√3) R30°, (4x4), and (√13x√13) R13.9°. These ordered phases, obtained by varying the substrate temperature during the silicon growth, correspond simply to different orientations of the silicene sheet relative to the Ag(111) surface[8,16]. Recently ARPES measurements performed on the silicene sheet on the (4x4) superstructure have reported a linear dispersion (Dirac cone) around the K-point of the Brillouin zone.[15]

So far STM images, which are dominated by electronic contrast, have formed the basis for the reported atomic models of silicene on Ag(111).[9,13-16] However, the atomic structure is still under intense debate. Non-contact atomic force microscopy (nc-AFM) measurements should permit us to discriminate between the electronic and geometric contributions and provide additional information enabling the construction of a realistic model for the structure of silicene adsorbed on Ag(111).

In this letter, we present combined nc-AFM/STM measurements using a qPlus sensor of the 4x4 superstructure of silicene grown on Ag(111) surface at room temperature (RT). We observe that constant height nc-AFM images show very similar atomic contrast to the STM topography. This similarity indicates that the observed atomic scale contrast is mainly a topographic effect instead of an electronic effect. This is unambiguous evidence that the investigated (4x4) silicene superstructure has a strongly buckled honeycomb structure.

All experiments and procedures were performed in an ultra-high vacuum (UHV) chamber of a modified variable temperature scanning probe microscope (Omicron VT XA qPlus nc-AFM/STM), which is capable of simultaneous measurements of the tunneling current and interaction forces. The heart of the microscope is the mechanically resonated, qPlus sensor[17] with a tungsten tip mounted at the extremity of the free prong. This particular sensor has a resonant frequency of 51264 Hz and a stiffness of 2975 N/m, which was determined from a thermal noise analysis of the sensor[18]. The internal wiring of the microscope was modified in order to eliminate any cross-talk phenomena, which can occur between the deflection channel and the tunneling current[19]. A Specs-Nanonis OC4 PLL controller was used for the frequency demodulation and the Omicron MATRIX control system for the data acquisition. The base pressure of the chamber was always below $1x10^{-10}$ mbar during data acquisition and all results were recorded at room temperature.

The Ag(111) single crystal was cleaned in UHV by ion sputtering ($Ar^+$ 500~eV) and annealing to 700 K. Evaporation of silicon was achieved by passing a direct current through a $10x4x0.5$ $mm^3$ piece from a silicon wafer. A reasonably stable Si flux, approximately 0.5 monolayer (ML) per minute, was achieved this way. Si was deposited onto the clean Ag(111) surface held at 500 K. After Si deposition, the sample was transferred to the microscope in the same UHV chamber for characterization with the nc-AFM/STM. Constant current operation was chosen for STM imaging. On the other hand the nc-AFM measurements were performed using a slow feedback on the frequency shift signal to compensate the tilt between the plane of the surface and the scanning plane. Furthermore, this procedure allows the drift in Z direction to be eliminated, which can be quite significant and non-linear at room temperature. Only the constant separation between the tip and the sample was maintained by this semi-constant height mode, no atomic corrugation was apparent in the topographic (Z) channel.

Fig.1a shows an atomically resolved STM topography of a selected area on the sample, containing two known reconstructions;[13-16] the (√13x√13) R13.9° and the (4x4). As a result of the sample preparation, the entire sample is covered by large areas of the (4x4) structure and only small areas of the (√13x√13) R13.9° were obtained, so we focused on the (4x4) structure to obtain atomically resolved AFM images using the constant height nc-AFM mode. Interpreting the nc-AFM images is quite challenging since the presence of the tunneling current alters the observed frequency shift. A recent study[20] has shown that the contrast inversion is bias dependent, and at high voltage biases is governed by the sample resistance and tunnel current. Consequently, we chose to record all images at zero bias voltage, where the tunnel current is negligible, and the observed atomic contrast is interpreted as a direct consequence of the short-range chemical interactions between the tip and sample.

In order to align the STM and AFM images, the operation mode was changed during the scanning. This allows the atomic positions to be well matched. The resulting STM topography and nc-AFM image of the same area are shown on Fig.1b and Fig.1c respectively. The nc-AFM image, Fig.1c, was taken in the Z height where the repulsive force over the imaged Si atoms prevails. Therefore, the bright protrusions in the image correspond to sites that had a greater repulsive action on the tip apex. The STM and nc-AFM images in Fig.1b and Fig. 1c are remarkably similar, showing protrusions at identical locations in the unit cell. Each triangle represents one half of the elementary surface unit cell. For a more detailed understanding of the tip-sample interaction mechanism, constant height scans were performed at different tip-sample separations. Fig.2 shows three distinct conditions of nc-AFM contrast as the tip is approached towards the (4x4) structure. First the tip was positioned far from the sample with a feedback set point of -10.5 Hz. At this distance only the contour of the unit cells can be resolved as shown in Fig. 2a. The absence of atomic contrast is due to negligible short-range interactions. However, the internal part of the unit cell is apparently more attractive than the corners. To enhance the atomic contrast, we reduced the tip-sample separation by changing the set point to -14 Hz. We can distinguish six bright protrusions forming two characteristic edge-to-edge triangles as observed in the STM mode. Therefore we can attribute the protrusions to the center atoms of the (4x4) phase. Surprisingly the probe feels the most attractive interaction in middle of the (4x4) unit cell while the corner hollow and 6 center atoms show very similar magnitude of frequency shift (see Fig.2b). The attractive interaction observed over the central part of the unit cell can be attributed to uniformly distributed long-range van der Waals forces. On the other hand, the tip-sample interaction over the 6 center Si atoms already shows the compensating contribution of a repulsive short-range force. At the closest tip-sample distance with set point of -17 Hz, the topmost 6 center Si atoms of the (4x4) structure are clearly resolved (see Fig.2c). In Fig.2d a cross sectional cut is shown. It was taken from the frequency shift image in Fig.2c along the long diagonal of the unit cell (the arrow indicates the position and direction of the line profile). We see that the frequency shift over the topmost Si atoms is -16.5 Hz. In fact, this is about 1 Hz larger than the average shift over corner hollow sites.

Consequently, the topmost Si atoms are less attractive than hollow sites. Again this observation can be only interpreted by the presence of the repulsive short-range interactions over the topmost Si atoms. The lack of a strong attractive short-range force compared to other Si-based surfaces,[21,22] indicates that either our tip apex was not chemically active or the silicene (4x4) structure is chemically inert. The later could explain low reactivity towards molecular oxygen[11]. We should note that approaching the tip closer towards the surface leads to unstable scanning conditions.

The atomic contrast acquired by nc-AFM and STM consists of 6 characteristic protrusions arranged in 2 triangles within the unit cell. The observation coincides very well with the proposed (4x4) model structure[8,16] as can be seen directly in Fig.1d. In particular, the (4x4) model structure shows clearly that the Si atoms are relaxed out of the surface plane forming pyramids, whose apices are located at the positions where both the STM and AFM show bright protrusions. This structural relaxation breaks the symmetry of the honeycomb structure over a distance corresponding to three honeycomb unit cells.

In summary, we have investigated the electronic and topographic landscape of a silicene sheet grown on Ag(111) surface by means of simultaneous STM and nc-AFM techniques. The nc-AFM images are identical to the STM images but only when the tip is relatively close to the surface. This is explained by the contribution of repulsive short-range forces which are strongest at the atomic positions of the Si atoms that are nearest to the tip. For the (4x4) superstructure of silicene on Ag(111), the nc-AFM measurements clearly indicate a strong buckling of the Si honeycomb layer.

**Acknowledgement**

Z.M, M.Š, P.H. and P.J. acknowledge financial support of GAAV under grant no. M100101207. RT and HO acknowledge the financial support from the European Community FP7-ITN Marie-Curie Programme (LASSIE project, grant agreement #238258) and the help of Prof. J.L. Lemaire its scientific coordinator in France at the Observatoire de Paris. AK thanks the ISMO group for hospitality.

**Images:**

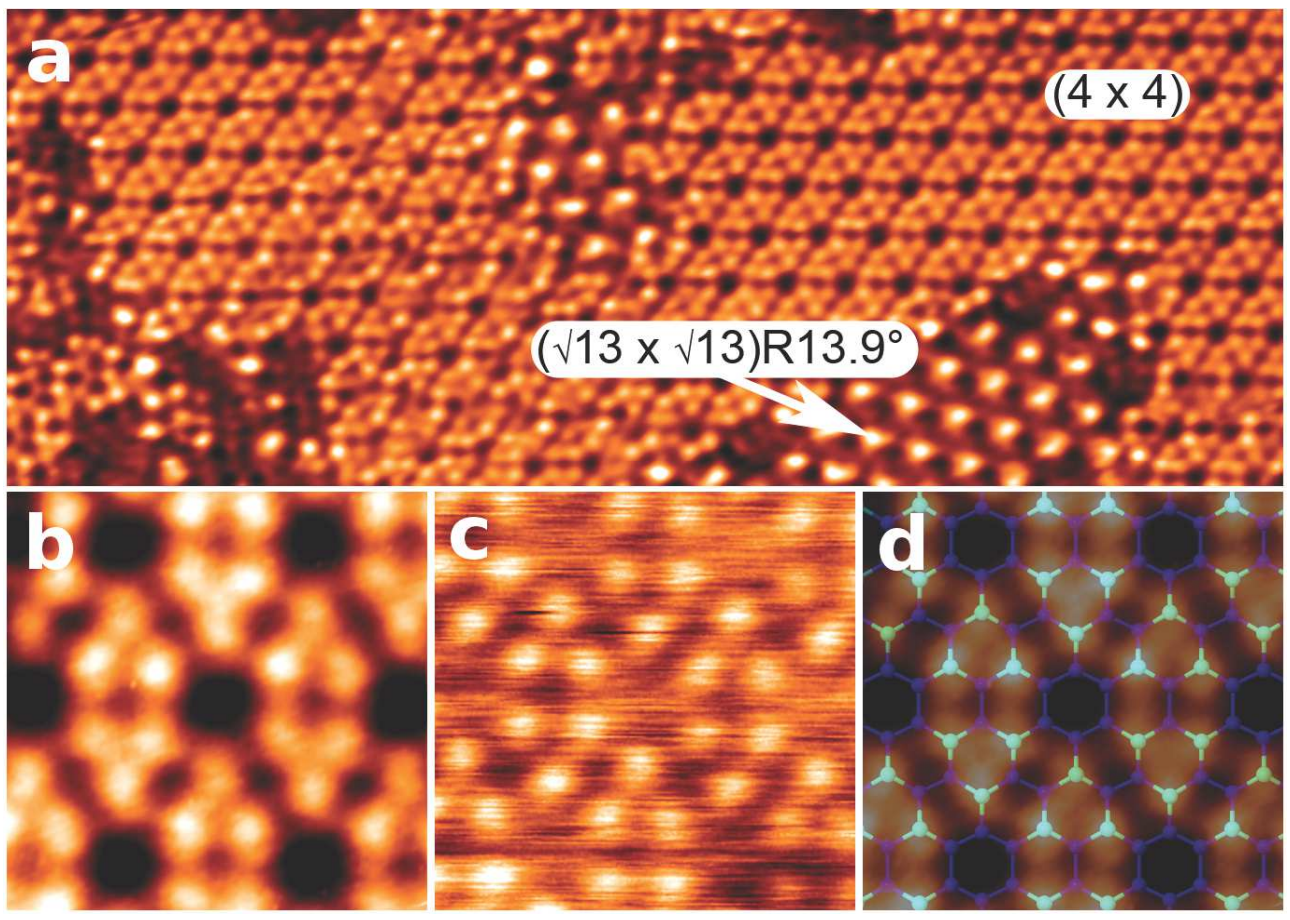

Fig.1: Atomically resolved STM topography (size) of the Ag(111) sample after 7 min deposition of Si at 500 K (a). The selected area on the sample, that contains two known reconstructions identified as (√13x√13)R13.9° and (4x4). $U_{BIAS}$, $I_t$. Image (b) is the result of a zoomed STM scan on the previously presented (4x4) area, with identical scanning condition. The same area was mapped in constant height nc-AFM operation as well (c). For the latter imaging -17 Hz was used as a feedback set point. (d) The proposed model for the (4x4) superstructure [8,16] matches very well our data.

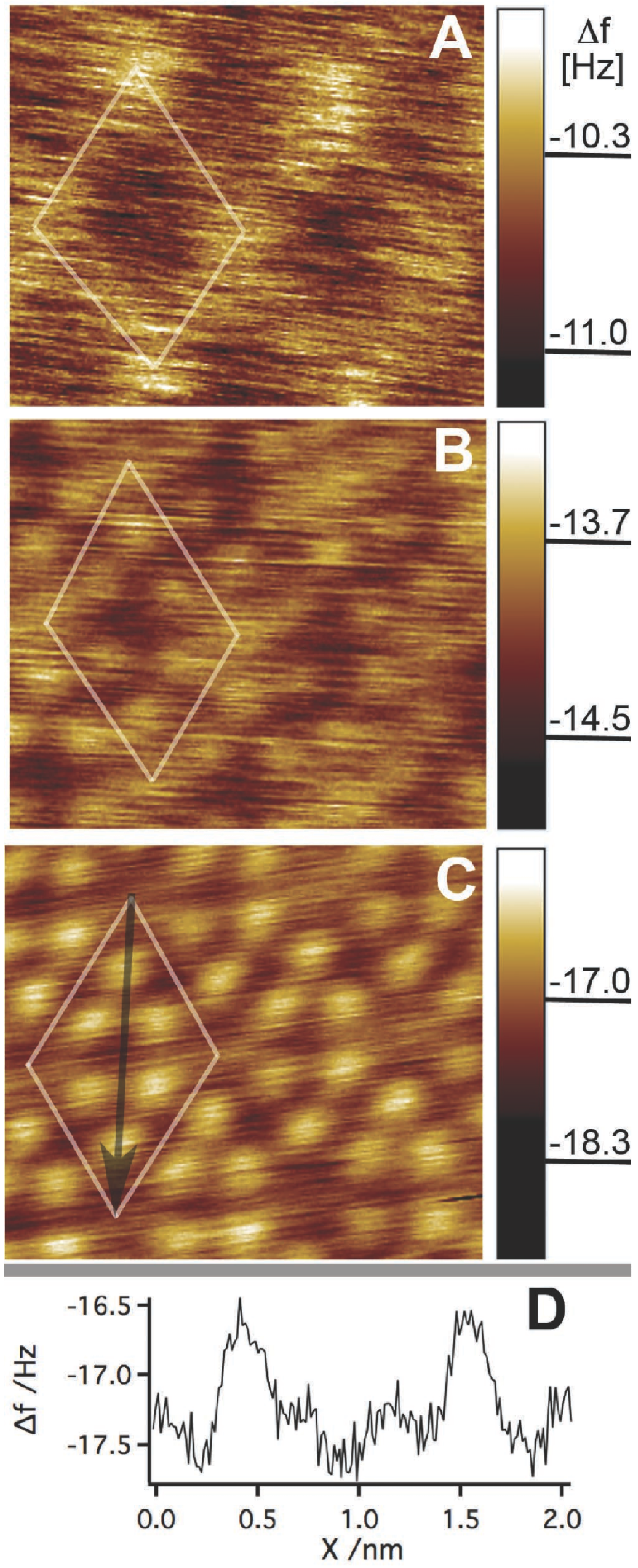

Fig.2.: Constant height, nc-AFM, which was maintained by a slow feedback regulated on Δf, scans of the same area. The tip-sample separation was gradually decreased by adjusting the Δf set point to: -10.5 (a), -14.0 (b) and -17 Hz (c). The size of images is 3 x 2.6 nm$^2$ and they are collected with $U_{BIAS}$ = 0.0 V. Fig (d) shows line profile of frequency shift Δf obtained along direction indicated by a arrow shown on image (c).